\documentclass[aps,twocolumn,nofootinbib,preprintnumbers,groupedaddress,letterpaper]{revtex4}

\usepackage[dvips]{color}
\usepackage[normalem]{ulem}
\usepackage{amsmath}
\usepackage{enumerate}
\usepackage{amsfonts}
\usepackage{epsfig}
\usepackage{yfonts}
\usepackage{undertilde}

\usepackage{graphicx}
\usepackage{bm}
\usepackage{subfigure}

\newcommand\comment[1]{}

\newcommand\poincare{Poincar\' e }

\newcommand\ov{\over }

\def\le{\left}
\def\ri{\right}

\def\({\left(}
\def\){\right)}
\def\[{\left[}
\def\]{\right]}
\def\<{\langle}
\def\>{\rangle}

\newcommand\half{{\ensuremath{\frac{1}{2}}}}
\newcommand\p{\ensuremath{\partial}}

\newcommand\field[1]{{\ensuremath{\mathbb{{#1}}}}}

\newcommand{\RR}{\field{R}}

\newcommand{\be}{\begin{equation}}
\newcommand{\ee}{\end{equation}}
\newcommand{\bea}{\begin{eqnarray}}
\newcommand{\eea}{\end{eqnarray}}
\newcommand{\bwt}{\begin{widetext}}
\newcommand{\ewt}{\end{widetext}}

\newcommand{\bi}{\begin{itemize}}
\newcommand{\ei}{\end{itemize}}
\newcommand{\ben}{\begin{enumerate}}
\newcommand{\een}{\end{enumerate}}
\newcommand{\bca}{\begin{cases}}
\newcommand{\eca}{\end{cases}}
\newcommand{\bln}{\begin{align}}
\newcommand{\eln}{\end{align}}
\newcommand{\bst}{\begin{split}}
\newcommand{\est}{\end{split}}

\begin{document}

\title{Segmented strings coupled to a B-field}

\author{David Vegh}
\email{dvegh@cmsa.fas.harvard.edu}

 \affiliation{\it  CMSA, Harvard University, Cambridge, MA 02138, USA   }

\date{\today}

\begin{abstract}

In this paper we study segmented strings in AdS$_3$ coupled to a background two-form whose field strength is proportional to the volume form. By changing the coupling, the theory interpolates between the Nambu-Goto string and the $SL(2, \RR)$ Wess-Zumino-Witten model. In terms of the kink momentum vectors, the action is independent of the coupling and the classical theory reduces to  a single discrete-time Toda-type theory. The WZW model is a singular point in coupling space where the map into Toda variables degenerates.

\end{abstract}

\maketitle

\section{Introduction}

Segmented strings in flat space are piecewise linear classical string solutions: at a given time their shape is a union of straight string segments. Kinks between the segments move with the speed of light and their worldlines form a lattice on the worldsheet.  This idea can be generalized to AdS$_3$ (or more generally to (A)dS$_n$) where the embedding is built from AdS$_2$ patches \cite{Vegh:2015ska, Callebaut:2015fsa}. The construction provides an exact discretization of the non-linear  string equations of motion (as opposed to approximate discretizations used for instance in \cite{Ishii:2015wua}).
For  recent developments, the reader is referred to \cite{Vegh:2015yua, Gubser:2016wno, Gubser:2016zyw, Ficnar:2013wba}.

In \cite{Vegh:2016hwq}, the area of segmented strings has been computed using cross-ratios constructed from the kink momentum vectors. The cross-ratios were expressed in terms of purely left-handed (or right-handed) Toda variables.
In this way, classical Nambu-Goto string theory in AdS$_3$ could be reduced to an integrable time-discretized relativistic Toda-type lattice.

In \cite{Gubser:2016wno}, Gubser pointed out that the segmented string evolution equations simplify if they are derived from the $SL(2, \RR)$ Wess-Zumino-Witten action. In this theory, strings couple to the three-form field strength which supports the AdS$_3$ geometry. (For the quantum theory see \cite{Maldacena:2000hw}.) If the three-form field strength comes from the Ramond-Ramond sector, then the WZW theory describes the motion of D1-branes, whereas the Nambu-Goto action is appropriate for fundamental strings (F1).

In this paper we generalize \cite{Vegh:2016hwq} by considering F1-D1 bound states.
Our starting point is the action \cite{Gubser:2016zyw}
\be
  \label{eq:action}
  S  = -{\tau_1 \ov 2}\int d^2 \sigma \, \p_a Y^M \p_b Y^N (\sqrt{-h}h^{ab} G_{MN} + \kappa \epsilon^{ab} B_{MN})
\ee
where $\tau_1$ is the tension. In order to simplify the formulas, we set the prefactor to one ($\tau_1= -2$). $Y^M$ are coordinates on AdS$_3$, $h$ and $G$ are the worldsheet and background metrics, respectively. $B$ is the background two-form with field strength proportional to the volume form of AdS$_3$.
Finally, $\kappa$ is its coupling to the worldsheet.

\begin{figure}[h]
\begin{center}
\includegraphics[scale=0.5]{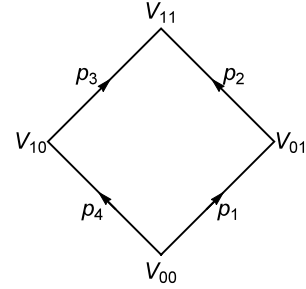}
\caption{\label{fig:momcon} A single worldsheet patch. The four edges are the kink worldlines where the normal vector jumps. In $\RR^{2,2}$ these are straight null lines with direction vectors $p_i$.
}
\end{center}
\end{figure}

\noindent
On causal grounds \cite{Gubser:2016zyw}, its value is restricted to $\kappa \in [-1,1]$. We can without loss of generality restrict $\kappa$ to be non-negative.

The equations of motion derived from (\ref{eq:action}) are exactly solved by segmented strings. These can be built by gluing diamond-shaped worldsheet patches. Each patch borders four others along null kink lines.

Let us consider the patch in FIG. \ref{fig:momcon}. The four vertices in the $\RR^{2,2}$ embedding space of AdS$_3$ are labeled by $V_{ij}$. We have $ V_{ij}^2 = -1$. The boundary of the worldsheet patch consists of four null kink lines.
Let us define the following kink momentum vectors
\bea
  \nonumber
   p_1 =  V_{01} -  V_{00} &\qquad &
   p_2 =  V_{11} -  V_{01} \\
  \label{eq:difvec}
   p_3 =  V_{11} -  V_{10} &\qquad &
   p_4 =  V_{10} -  V_{00}
\eea

These vectors satisfy
\be
  \nonumber
   p_i^2 = 0  \qquad \textrm{and} \qquad   p_1 +  p_2  =  p_3 +  p_4
\ee

The latter equation can be interpreted as ``momentum conservation'' during the scattering of two massless scalar particles with initial and final momenta $ p_{1,2}$ and $ p_{3,4}$, respectively.

Let $X(\sigma^-, \sigma^+) \in \RR^{2,2} $ denote the embedding function of the string into spacetime where $\sigma^\pm$ are lightcone coordinates on the worldsheet.
The patch is bounded by
\bea
  \nonumber
  X(\sigma^-, 0) &=& V_{00} + \sigma^- p_4  \\
  \nonumber
  X(0, \sigma^+) &=& V_{00} + \sigma^+ p_1
\eea
for $\sigma^\pm \in (0,1)$.

Points on the surface are given by the {\it interpolation ansatz} \cite{Gubser:2016zyw} which solves the equations of motion
\bea
  \nonumber
  X (\sigma^-, \sigma^+) &=&  {1+ (1+\kappa^2)\sigma^- \sigma^+ p_4 \cdot p_1 /2 \over
  1 - (1-\kappa^2)\sigma^- \sigma^+ p_4 \cdot p_1 /2} V_{00} + \\
    &+& {\sigma^-p_4 + \sigma^+ p_1 + \kappa \sigma^+\sigma^- N \over 1-(1-\kappa^2)\sigma^+\sigma^-  p_4 \cdot p_1 /2}
  \label{eq:ian}
\eea
where 
\be
  \nonumber
  N^\mu = \epsilon^\mu_{\nu\lambda\rho} V^\nu_{00} p_4^\lambda p_1^\rho
\ee
From the interpolation ansatz, we have
\be
  V_{11} = X(1,1) .
\ee
This equality constitutes a discrete evolution equation for segmented strings. The set of points in AdS$_3$ null separated from both $V_{10}$ and $V_{01}$ is a one-dimensional locus, conveniently parametrized by $V_{11}(\kappa)$. Note that at $\kappa=0$, the interpolation ansatz gives an AdS$_2$ patch embedded into AdS$_3$.

In the next section, we evaluate the action for a single patch. The action can be written in terms of Mandelstam variables corresponding to kink momentum vectors. Section III computes the Euler character for the string. Section IV discusses the singular $\kappa=1$ case where the map into Toda variables degenerates.
We finish with a short discussion of the results.

\section{The action of segmented strings }

The value of the action evaluated on the patch is analogous to a scattering amplitude in four dimensional Minkowski spacetime. Instead of the Lorentz symmetry, however, the action is invariant under the $SO(2,2)$ isometry group of AdS$_3$ and $\RR^{2,2}$.  The only independent invariants are the Mandelstam variables $s = ( p_1+  p_2)^2$ and  $u= ( p_1 -  p_4)^2$. The patch action then takes the form
\be
  \nonumber
  S_\textrm{patch} = L^2 \mathcal{F}\le( {u \over s} \ri)
\ee
where $L$ is the AdS$_3$ radius (henceforth set to one) and $\mathcal{F}(x)$ is a dimensionless function. In the following, we will determine this function by evaluating the action for patches that have been moved into a certain position.

\subsection{Fixing the patch}

Using the $SO(2,2)$ symmetry we can rotate/boost any patch such that $V_{00},V_{10},V_{01}$ are parametrized by two numbers $c$ and $\tilde c$ as follows
\bea
  \nonumber
   V_{00} &=& (1, \, 0,  \,\,\,\, 0,   \, 0)^T \\
  \nonumber
   V_{10} &=& (1,\,c,-c, \, 0)^T \\
  \nonumber
   V_{01} &=& (1,\,\tilde c, \,\,\,\,\, \tilde c, \, 0)^T .
\eea

These points satisfy $ V_{ij}^2 = -1$.
It is easy to check that the corresponding difference vectors from (\ref{eq:difvec}) indeed
satisfy $ p_i^2 = 0$.

The interpolation ansatz in (\ref{eq:ian}) gives the embedding of the patch into $\RR^{2,2}$
{\footnotesize
\be
  \label{eq:inpol}
 X(\sigma^{-},\sigma^{+}) =
 {1\ov 1+c\tilde c(1-\kappa^2)\sigma^{-} \sigma^{+}} \left(
\begin{array}{c}
   1- c\tilde c \left(1+\kappa ^2\right) \sigma^{-} \sigma^{+}  \\
   c \sigma^{-}+\tilde c \sigma^{+} \\
   -c \sigma^{-}+\tilde c \sigma^{+} \\
  -2 c\tilde c \kappa  \sigma^{-} \sigma^{+}
\end{array}
\right)
\ee
}
The fourth vertex is computed from $V_{11} = X(1,1)$.
\comment{
\be
  \nonumber
 V_{11} = {1\ov 1+ab(1-\kappa^2)} \left(
\begin{array}{c}
 1-ab(1+\kappa^2) \\
  a+b \\
    b-a \\
   -2ab\kappa
\end{array}
\right)
\ee
}

The Mandelstam variables are found to be
\be
  \nonumber
  s = -{4 c \tilde c \over 1+c\tilde c(1-\kappa^2)} \qquad  \textrm{and} \qquad  u = 4 c \tilde c
\ee
from which
\be
  \label{eq:x}
  {u\over -s} = 1+c\tilde c(1-\kappa^2)
\ee
for the argument of $\mathcal{F}(x)$. Note that $\kappa=1$ is a special point, since at this value the ratio is independent of $c\tilde c$. This is precisely the Wess-Zumino-Witten theory.  

\subsection{Evaluating the action}

Let us now evaluate action (\ref{eq:action}) for our worldsheet patch. The metric and a canonical B-field are
\be
  \nonumber
  ds^2 = {-dt^2 + dz^2 \ov z^2} \qquad B_0 =  \frac{dx \wedge dt}{2 z^2} \, .
\ee
The field strength three-form is
\be
  \nonumber
  H = dB_0 = -\textrm{Vol}_{\textrm{AdS}_3}
\ee
Since $H$ and the background metric are both $SO(2,2)$ invariant, we expect that the patch action will also be invariant and therefore it can be expressed in terms of Mandelstam variables.

The $\kappa=0$ case is straightforward and the results were presented in \cite{Vegh:2016hwq}. At $\kappa\ne 0$, however, $S(B_0)$ depends separately on $c$ and $\tilde c$ and thus it cannot be expressed in terms of Mandelstam variables  which are functions of $c \tilde c$. This is due to the fact that  the bulk action alone is not gauge-invariant if the worldsheet has boundaries. In order to preserve gauge invariance, point particles with opposite charges must be attached to the string endpoints. These particles are charged under a one-form gauge field $A$. The minimal coupling is described by
\be
  \nonumber
  S_A = \int_{\p \Sigma} A
\ee
where $\p \Sigma$ is the worldsheet boundary.
The variation of the bulk worldsheet action under a gauge transformation $\Lambda$ can be canceled if we let $A$ transform according to
\bea
  \nonumber
  B &\to& B + d\Lambda \\
  \nonumber
  A &\to& A + \Lambda
\eea
Our strategy will be the following. For a given $B$, we choose $A$ such that $S+S_A$ is $SO(2,2)$ invariant (and thus a function of $s,t,u$). Clearly, $A$~does not affect closed string motion because those worldsheets have no boundaries.  Using the gauge transformation above, we set $A=0$. As a result, $S_A$ vanishes and the entire patch action will come from  $S$.
When the dust settles, all we have done was a gauge transformation on $B$ (and we can forget about $S_A$).

Consider the following B-field
\bea
  \nonumber
  B &=& {b_+ \over 2z^2 b_-} dx \wedge dt  +  {x \over z b_-} dt \wedge dz + {1-t \over z b_-} dx \wedge dz \\
  \nonumber
& &  \textrm{with} \quad  b_\pm \equiv 1+t^2-2 t-x^2 \pm z^2
\eea
It is gauge-equivalent to $B_0$ and -- as we will see -- gives an $SO(2,2)$ invariant result.

Let us now evaluate (\ref{eq:action}) on the patch given by the interpolation ansatz (\ref{eq:inpol}) using the expression above  for the B-field. The purely geometrical part of the integrand gives
\be
  \label{eq:i1}
  I_1 = \frac{2 c\tilde c}{\left(c\tilde c \left(\kappa ^2-1\right)\sigma^{-} \sigma^{+}-1\right)^2}
\ee
The second term (proportional to the B-field) gives
\be
  \label{eq:i2}
  I_2 = -{\kappa^2  } I_1
\ee
Integrating $I_1 + I_2$ over the patch gives the value of the action. We get
\be
  \nonumber
  S_\textrm{patch} =  \int_0^1 d\sigma^-\int_0^1 d\sigma^+ (I_1 + I_2) = 2 \log[1+c\tilde c(1-\kappa^2)]
\ee
Combining these results with (\ref{eq:x}) results in the covariant formula
\be
  \label{eq:covarea}
   S_\textrm{patch} = 2 \log   {u\ov -s}
\ee
Note that the expression is independent of $\kappa$.

\subsection{Toda variables}

The Mandelstam variables may be expressed in terms of Toda variables as in \cite{Vegh:2016hwq}.
In order to do this, the lightlike kink momenta $p$ are written as products of helicity spinors.
We define
\bea
  \nonumber
  \sigma^\mu &=& ( 1, -i \sigma_2, \sigma_1, \sigma_3 ) \\
  \nonumber
   p_{a\dot a} &=& \sigma^\mu_{a\dot a} p_\mu
\eea
Since $p^2 = \det(p_{a\dot a}) = 0$, we can write
\be
  \nonumber
  p_{a\dot a} = \lambda_a \tilde\lambda_{\dot a}
\ee

\begin{figure}[h]
\begin{center}
\includegraphics[scale=0.45]{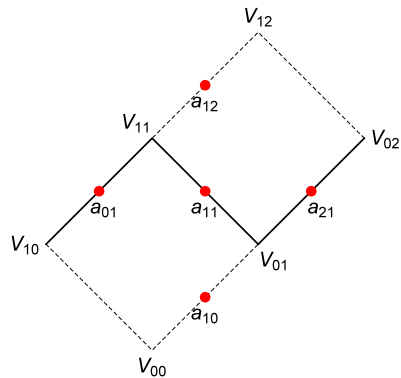}
\caption{\label{fig:double}  Two adjacent patches on the worldsheet. $V_{00}$ and $V_{12}$ can be computed using the interpolation ansatz. Then, the five left-handed Toda variables $a_{ij}$ computed from the difference vectors will satisfy the Toda-type equation of motion.
}
\end{center}
\end{figure}

We will call $\lambda$ left-handed spinors and $\tilde\lambda$ right-handed spinors. In terms of these two-component variables, the patch action can be written as
\be
  \nonumber
   S_\textrm{patch} = 2 \log\le|  {\langle\lambda_1, \lambda_4 \rangle\langle\lambda_2, \lambda_3 \rangle
  \over \langle\lambda_1, \lambda_2 \rangle\langle\lambda_3, \lambda_4 \rangle  } \ri|
\ee
There is a similar formula in terms of  right-handed spinors.
The spinor modulus drops out of the action. Thus, by defining the angles $\alpha_i$ via
\be
  \nonumber
  |\lambda_i| e^{i\alpha_i} := \lambda_i^1 + i\lambda_i^2
\ee
one can write
\be
   \nonumber
 S_\textrm{patch} =  2\log\le|   {\sin(\alpha_{1} - \alpha_{4})\sin(\alpha_{2} - \alpha_{3}) \over \sin(\alpha_{1} - \alpha_{2})\sin(\alpha_3 - \alpha_{4})} \ri|
\ee
The $\alpha$ angles are the {\it global} Toda variables.
Let us further define the left-handed {\it \poincare} Toda variables  by
\be
  \nonumber
  a_i := \tan \alpha_i \, .
\ee
For a kink momentum vector $p \in \RR^{2,2}$, the left-handed and right-handed \poincare Toda variables are simply given by
\be
  \label{eq:ap}
  a(p) = {p_{-1} + p_2 \over  p_0 + p_1}, \qquad \tilde a(p) = {p_{-1} + p_2 \over -p_0 + p_1}
\ee
In terms of these fields the patch action becomes
\be
  \label{eq:patcharea}
  S_\textrm{patch} =   2 \log\le| {(a_1-a_4)(a_2-a_3) \ov (a_1-a_2)(a_3-a_4) }\ri| \, .
\ee
The total action is then the sum of all patch contributions
\be
  \label{eq:aaction}
  S = 2 \sum_{i,j}   \log\le| a_{i,j}  - a_{i+1,j}   \over  a_{i,j}  - a_{i,j+1}  \ri|
\ee
where the $i,j$ indices indicate the position of the kink edge in the two-dimensional lattice (see FIG. \ref{fig:sub}).
The previously used one-index $a_k$ are
\bea
  \nonumber
   a_1 &\rightarrow& a_{ij} \qquad \qquad  \,  a_2 \rightarrow a_{i,j+1}  \\
  \nonumber
   a_3 &\rightarrow& a_{i-1,j+1}  \qquad   a_4 \rightarrow a_{i-1,j}
\eea
and $a_{ij}$ sits on a white dot in the lattice, see FIG. \ref{fig:sub}.

\subsection{Equation of motion}

\label{sec:eom}

The equation of motion computed from (\ref{eq:aaction}) is \cite{Vegh:2016hwq}
\be
  \nonumber
  \hskip -0.1cm {1\ov a_{i,j} - a_{i,j+1}}+   {1\ov a_{i,j} - a_{i,j-1}} =
  {1\ov a_{i,j} - a_{i+1,j}}+   {1\ov a_{i,j} - a_{i-1,j}}
\ee
This equation is independent of $\kappa$. It has been obtained in \cite{suris} as the equation of motion of a time discretization of a relativistic Toda-type lattice.

The validity of the above equation of motion can be checked directly. Let us consider two adjacent patches as in FIG. \ref{fig:double} and fix four vertices $V_{10}$, $V_{11}$, $V_{01}$, $V_{02}$ in $\RR^{2,2}$ such that neighboring vertices are lightlike separated. The interpolation ansatz (\ref{eq:ian}) then gives\footnote{Note that in order to get $V_{00}$ from  $V_{10}$, $V_{11}$, $V_{01}$, the sign of $\kappa$ must be reversed in the formula.}  $V_{00}$ and $V_{12}$. The left-handed Toda variables $a_{ij}$ can be computed from the kink momentum vectors. We claim that these variables satisfy the above equation of motion.

The calculation is straightforward using a computer algebra program, but the formulas are too large to present here. Therefore we only sketch the proof. One might start with the following ``forward null triple''
\bea
  \nonumber
   V_{01} &=& (1, \, 0,  \, 0,   \, 0)^T \\
  \nonumber
   V_{11} &=& (1, \sqrt{c_1^2+c_2^2}, \, c_1, \, c_2)^T \\
  \nonumber
   V_{02} &=& (1, \sqrt{c_3^2+c_4^2}, \, c_3, \, c_4)^T .
\eea
supplemented by another vertex $V_{10}$ which satisfies
\be
  \nonumber
  V_{10}^2 = -1 \qquad \textrm{and} \qquad (V_{10}-V_{11})^2 = 0 \, .
\ee
Altogether there are six real constants parametrizing the vertices. Note that $V_{01}$ is fixed, but apart from this the ansatz is generic.  In principle, one could start with a completely generic null triple ansatz, but then the formulas become prohibitively complicated. Instead, we fix $V_{01}$, and we perform a global $SO(2,2) = SL(2)_L \times SL(2)_R$ transformation on the kink momentum vectors before the final step when the Toda variables are computed. Note that left-handed variables are invariant under $SL(2)_R$ and thus it is enough to consider $SL(2)_L$ rotations.
The resulting Toda variables do satisfy the equation of motion.

\section{Euler character}

In this section, for pedagogical reasons, we study the Euler character
\be
   \nonumber
  \chi = {1\ov 4\pi } \int d^2\sigma \sqrt{-g} \, R \, .
\ee
Let us consider a worldsheet with torus topology\footnote{This is possible if the target space is the surface $X^2 = -1$, without going to the covering space. Equivalently, one may consider time-periodic solutions in AdS$_3$.}. Such worldsheets are known to give $\chi=0$. In the following, we will check this result for segmented strings.

Using the interpolation ansatz (\ref{eq:ian}) and $\sigma^\pm = \half(\tau\pm \sigma)$, the induced metric is diagonal with components
\be
   \nonumber
  g_{\sigma\sigma} = -g_{\tau\tau}=\frac{ c\tilde c}{\left(1+ {1-\kappa ^2 \ov 4}  \left(\tau ^2-\sigma ^2\right)c\tilde c\right)^2} \, .
\ee
The Ricci scalar computed from $g$ is constant away from kink collision vertices
\be
  \label{eq:rcc}
  R_0 = -2(1-\kappa^2)
\ee
Let us decompose the Euler character as an integral away from the vertices plus integrals at the vertices
\be
   \label{eq:euler}
   4\pi \chi =  R_0 \int d^2\sigma \sqrt{-g} + \sum_i \int_{\mathcal{V}_i} d^2\sigma \sqrt{-g} \, R  \, .
\ee
Here the sum is over kink vertices and $\mathcal{V}_i$ labels an infinitesimally small area around the $i^\textrm{th}$ vertex.

Let us momentarily set $\kappa=0$. From the appendix of \cite{Vegh:2016hwq}, we have
\be
  \nonumber
  \int_{\mathcal{V}_i} d^2\sigma \sqrt{-g} \, R \stackrel{\kappa=0}{=}   8 \log \cos {\phi_i \ov 2}
\ee
where $\phi_i$ is defined such that $\cos \phi_i$  is the scalar product of the normal vectors of the two space-like separated patches around the $i^\textrm{th}$ collision point.

Denote the kink collision point by $X_M \in \RR^{2,2}$. Then, two kink momentum vectors emanating from this point span an AdS$_2$ patch with normal vector \cite{Vegh:2016hwq}
\bea
  \nonumber
   & N(\alpha,\beta) = {1\ov \sin(\alpha-\beta)} \times \hskip 6cm  &  \\
  \nonumber
  &
  \hskip -0.8cm  \times \left( \begin{array}{l}
   \, \, \, \, \, \, \, \, X_0 \cos(\alpha-\beta) - X_1 \cos(\alpha+\beta)  - X_2 \sin(\alpha+\beta)  \\
   -X_{-1} \cos(\alpha-\beta) - X_2 \cos(\alpha+\beta)  + X_1 \sin(\alpha+\beta)  \\
   -X_2 \cos(\alpha-\beta) - X_{-1} \cos(\alpha+\beta)  + X_0 \sin(\alpha+\beta)  \\
    \, \, \, \, \, X_1 \cos(\alpha-\beta) - X_0 \cos(\alpha+\beta)  - X_{-1} \sin(\alpha+\beta)
\end{array} \right)
   &
\eea
Here $\alpha$ and $\beta$ are the left-handed global Toda variables corresponding to the two kink momentum vectors. Using this formula, $\phi_i$ can be expressed and we arrive at
\be
  \label{eq:ress}
  \int_{\mathcal{V}_i} d^2\sigma \sqrt{-g} \, R  
     \stackrel{\kappa=0}{=} 4\log\le| {(a^{(i)}_1-a^{(i)}_4)(a^{(i)}_2-a^{(i)}_3) \ov (a^{(i)}_1-a^{(i)}_2)(a^{(i)}_3-a^{(i)}_4) }\ri|
\ee
where $a^{(i)}_k$ are the four left-handed \poincare Toda variables around the  $i^\textrm{th}$ collision point on the worldsheet.

The result (\ref{eq:ress}) holds even for $\kappa\ne 0$. Recall that in \cite{Vegh:2016hwq}, the integrated Ricci scalar was computed in flat background space and then the result was expressed in terms of AdS$_3$ quantities (i.e. $\RR^{2,2}$ normal vectors). The curvature of the target space did not matter since the collision of kinks was instantaneous. The Christoffel symbols can therefore be neglected in a limit where we zoom in on the collision point. Similary, we can argue that the three-form field strength can also be neglected in this limit and thus (\ref{eq:ress}) should be independent of $\kappa$.

By performing the sum over kink collisions we get
\be
  \label{eq:stot}
  \sum_i \int_{\mathcal{V}_i} d^2\sigma \sqrt{-g} \, R = 
   4 \sum_{i,j}   \log\le|   a_{i,j}  - a_{i+1,j}  \over a_{i,j}  - a_{i,j+1} \ri|
\ee
where now the indices of $a_{ij}$ label positions in the kink lattice, see FIG. \ref{fig:sub}.
In the RHS, using the results of the previous section, we recognize twice the total action (\ref{eq:aaction}).

Plugging (\ref{eq:stot}) into (\ref{eq:euler}) and using (\ref{eq:rcc}) we  get
\be
  \label{eq:res}
  4\pi \chi =  -2(1-\kappa^2) A + 2S_\textrm{total}
\ee
where $A$ is the worldsheet area. From (\ref{eq:i1}) and (\ref{eq:i2}) the integrand in the bulk action is
\be
  \nonumber
  I_\textrm{total} = I_1 + I_2 = (1-\kappa^2)I_1
\ee
which, after integration, yields $ S_\textrm{total} = (1-\kappa^2) A$. Plugging this result back into (\ref{eq:res}) finally gives $\chi=0$.

\section{Degenerate WZW limit}

Classical solutions for the $SL(2)$ Wess-Zumino-Witten model are given by
\be
  \nonumber
  g = g_+(\sigma^+)g_-(\sigma^-) \in SL(2, \RR)
\ee
How can such a (classically) trivial theory be mapped into the non-trivial Toda-type theory?
The answer is that at $\kappa = 1$ the map is not surjective: segmented strings are mapped into a smaller subspace of the Toda phase space\footnote{Note that for strings with $|\kappa|<1$, only {\it positive} Toda solutions play a role. These are the field configurations for which elementary patch areas (\ref{eq:patcharea}) are non-negative.}.
The fact that this point in coupling space is singular can already be seen from (\ref{eq:x}) that gives $u/s = -1$ which is independent of the $c$ and $\tilde c$ patch parameters.

\begin{figure}[h]
\begin{center}
\includegraphics[scale=0.65]{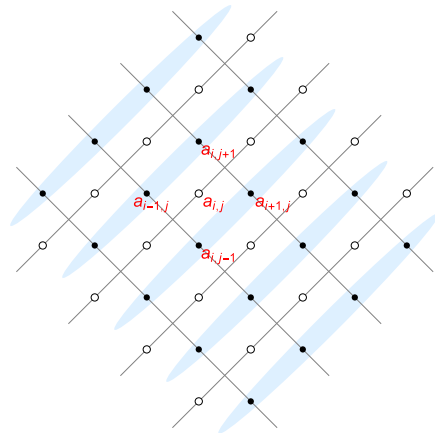}
\caption{\label{fig:sub} Kink worldlines form a rectangular lattice on the string worldsheet. The field $a_{ij}$ lives on the edges (black or white dots depending on edge orientation). For $\kappa=1$, the variables grouped together are equal (blue shading).
}
\end{center}
\end{figure}

FIG. \ref{fig:sub} shows the trivial subspace of the left-handed Toda phase space.
The $a_{ij}$ variables sitting on the black dots depend only on $i+j$ and they are independent of $i-j$. This means that the black dots grouped together (blue shading) have the same values. This is clearly a lower dimensional subspace.
We have not included a separate (mirror) figure, but the right-handed variables $\tilde a_{ij}$ are similarly degenerate: the ones sitting on white dots depend only on $i-j$ (and not on $i+j$).

Let us sketch the proof of degeneracy discussed above. The calculation is similar to the one in section \ref{sec:eom}.

Consider the following forward null triple
\bea
  \nonumber
   V_{00} &=& (1, \, 0,  \,  0,   \, 0)^T \\
  \nonumber
   V_{10} &=& \le(1,\,c_1,-c_2, \, \sqrt{c_1^2-c_2^2} \ri)^T \\
  \nonumber
   V_{01} &=& \le(1,\,\tilde c_1, \,\,\,\,\, \tilde c_2, \, \sqrt{\tilde c_1^2-\tilde c_2^2} \ri)^T
\eea

The interpolation ansatz then gives the fourth vertex,
{\footnotesize
\be
\nonumber
V_{11} = {1\ov C} \left(
\begin{array}{c}
 { {\kappa ^2+1 \ov 2} \left(\sqrt{(c_1^2-c_2^2)(\tilde c_1^2-\tilde c_2^2)}-{c_1} {\tilde c_1}-{c_2} {\tilde c_2}\right)+1}  \\
 {-\kappa  \left({\tilde c_2} \sqrt{c_1^2-c_2^2}+{c_2} \sqrt{{\tilde c_1}^2-{\tilde c_2}^2}\right)+{c_1}+{\tilde c_1}}  \\
 {-{\tilde c_1} \kappa  \sqrt{c_1^2-c_2^2}+{c_1} \kappa  \sqrt{{\tilde c_1}^2-{\tilde c_2}^2}-{c_2}+{\tilde c_2}} \\
 {\sqrt{c_1^2-c_2^2}  +\sqrt{{\tilde c_1}^2-{\tilde c_2}^2}-\kappa({c_1} {\tilde c_2}+{\tilde c_1} {c_2})}
\end{array}
\right)
\ee
}
\be
  \nonumber
  C = {\kappa ^2-1 \ov 2} \left(\sqrt{(c_1^2-c_2^2)(\tilde c_1^2-\tilde c_2^2)}-{c_1} {\tilde c_1}-{c_2} {\tilde c_2}\right)+1
\ee

We now perform a global
\be
  \nonumber
  \mathcal{R} \in SO(2,2) = SL(2)_L \times SL(2)_R
\ee
transformation on the kink momentum vectors which transforms $V_{00}$ into a generic position. Left-handed variables are invariant under $SL(2)_R$ and thus it is enough to consider $w \in SL(2)_L$ rotations.
The left-handed Toda variables computed from the difference vectors are
\bea
  \nonumber
  a_1 \equiv a( \mathcal{R}(V_{01}-V_{00})) &=&  \frac{w_{11} \sqrt{\tilde c_1-\tilde c_2}+w_{12} \sqrt{\tilde c_1+\tilde c_2}}{ w_{21} \sqrt{\tilde c_1-\tilde c_2}+   w_{22}\sqrt{\tilde c_1+\tilde c_2}}
  \\
  \nonumber
  a_4 \equiv a(\mathcal{R}(V_{10}-V_{00})) &=& \frac{w_{11} \sqrt{ c_1+ c_2}+w_{12} \sqrt{c_1- c_2}}{ w_{21} \sqrt{ c_1+  c_2}+   w_{22}\sqrt{ c_1- c_2}}
\eea
where $a(p)$  denotes the \poincare Toda variable corresponding to a kink momentum vector $p$, see eqn. (\ref{eq:ap}).

The other two variables $a_2$ and $a_3$ depend on $\kappa$. They are easy to compute, but the formulas are too large to present here. They satisfy
\bea
  \nonumber
  a_2 \equiv a(\mathcal{R}(V_{11}-V_{01})) &\stackrel{\kappa \to 1}{\longrightarrow} & a_4 \\
  \nonumber
  a_3 \equiv a(\mathcal{R}(V_{11}-V_{10})) &\stackrel{\kappa \to -1}{\longrightarrow} & a_1
\eea

There are degeneracies in the right-handed variables which can be proven in a similar fashion.

We finish this section with the following observation. Let us exchange black and white dots in the lattice. This duality exchanges patches and kink collision vertices and changes the string embedding. After the transformation, left-handed variables sitting on white dots along the same kink line will be equal. This configuration correspond to trivial left-moving kinks (edges with black dots), since they do not cause a time delay when they cross a right-moving kink. Thus, this string embedding only contains right-moving shockwaves and in the $\kappa=0$ case it is equivalent to Mikhailov's construction \cite{Mikhailov:2003er}.

\section{Discussion}

In this paper, we have computed the action of segmented strings in AdS$_3$. The worldsheet is coupled to a background two-form whose field strength is proportional to the volume form.
We have used the interpolation ansatz of \cite{Gubser:2016zyw} to parametrize elementary patches. Segmented string solutions are obtained by gluing the patches along null boundaries (kink lines).

The null kink momentum vectors in the embedding space $\RR^{2,2}$ can be decomposed using helicity spinors.  Then, the action can be expressed in terms of cross-ratios of the spinor angles. We have called both these angles and their tangents ``Toda variables''. Time evolution of segmented strings can be described by the evolution equation of a discrete-time Toda-type lattice. This equation was presented in section \ref{sec:eom}.

Interestingly, the final form of the action does not depend on the two-form coupling $\kappa$. Thus, the theory in terms of Toda variables treats the classical Nambu-Goto theory ($\kappa=0$) and the $SL(2)$ Wess-Zumino-Witten model ($\kappa=1$) on the same footing.
However, the latter theory is a special one, because the map from the segmented string into Toda variables degenerates as $\kappa\rightarrow 1$. This is seen in FIG. \ref{fig:sub}: the variables grouped together will become equal at $\kappa=1$.  By performing a duality that exchanges black and white dots in the lattice, WZW solutions can be mapped to string solutions with purely left- or purely right-moving kinks.

The results generalize those in \cite{Vegh:2016hwq} which can be obtained by setting $\kappa=0$.
We have not discussed the reconstruction of string embeddings from solutions of the Toda-like lattice. The procedure should be analogous to the $\kappa=0$ case.

An interesting question is how these ideas generalize to other spacetimes (e.g. dS$_n$ or AdS$_n$) and what kind of Toda-like theories one would get by a similar reduction. We leave this for future work.

\vspace{0.2in}   \centerline{\bf{Acknowledgments}} \vspace{0.2in}
This work was supported by the Center of Mathematical Sciences and Applications at Harvard University.
I would like to thank Daniel Harlow for comments on the manuscript.


\end{document}